\documentclass[smallextended]{svjour3}     
\smartqed  
\usepackage{amsfonts}
\usepackage{amsmath}
\usepackage{amssymb}
\usepackage{epsfig}
\usepackage{fullpage}
\usepackage{graphicx}
\usepackage{ifthen}
\usepackage{textcomp}
\journalname{GRG}

\newcommand{\rd}{\mathrm{d}}
\newcommand{\gb}{\bar{g}}
\newcommand{\xb}{\bar{X}}
\newcommand{\f}{{f\left(r,\varphi\right)}}

\begin{document}

\title{Gravitational lensing in the Kerr-Randers optical geometry}
\titlerunning{Kerr-Randers optical geometry}
\author{M. C. Werner}
\authorrunning{Werner}
\institute{M. C. Werner \at Kavli Institute for the Physics and Mathematics of the Universe,
University of Tokyo, 5-1-5 Kashiwanoha, \\ Kashiwa 277-8583, Japan. \email{marcus.werner@ipmu.jp} 
\\ Department of Mathematics, Duke University, Durham, NC 27708, USA}
\date{Received: date / Accepted: date}
\maketitle

\begin{abstract}
A new geometric method to determine the deflection of light in the equatorial plane of the Kerr solution 
is presented, whose optical geometry is a surface with a Finsler metric of Randers type. Applying the 
Gauss-Bonnet theorem to a suitable osculating Riemannian manifold, adapted from a construction by Naz\i m, 
it is shown explicitly how the two leading terms of the asymptotic deflection angle of gravitational 
lensing can be found in this way.
\end{abstract}

\section{Introduction}
This year sees the centenary of Einstein's first calculations of the lensing effect due to the gravitational 
deflection of light (for a discussion of the history see \cite{s08}), which has grown into an expanding research field in 
mathematical physics as well as in cosmology and galactic astronomy where it holds, for instance, the possibility 
of measuring a Kerr black hole's angular momentum parameter. A comprehensive survey is given by a recent 
special issue of \textit{General Relativity and Gravitation} dedicated to this topic, including a review of lensing by black 
holes \cite{b10} and mathematical aspects of the theory \cite{pw10}.

In this article, we will revisit lensing in the equatorial plane of the Kerr solution and suggest a new way to evaluate the 
deflection angle, assuming that both light source and observer are in the asymptotically flat region and that the deflection is
small which is, of course, a common situation in astronomy. Two methods have been used so far to analytically find deflection angles 
in this limit: firstly, by directly computing the null geodesics of the Kerr spacetime and evaluating the constants of motion at 
the source and the observer (see \cite{akp11} for a recent comprehensive treatment, and references therein); and secondly, by considering 
the Kerr lens on the optical axis as a notional Schwarzschild lens whose displacement from the optical axis is proportional to the angular 
momentum parameter. This uses the standard quasi-Newtonian impulse approximation and is sufficient to determine the leading order 
dependence of gravitational lensing observables on the angular momentum parameter (see, e.g., \cite{aky03,wp07}). The method proposed here 
extends a suggestion by Gibbons and the author \cite{gw08} using optical geometry (also called optical reference geometry or Fermat geometry). In this approach to spherically 
symmetric lenses, a Riemannian metric manifold is considered whose geodesics are the spatial light rays. Given that a light ray is in a 
geodesically complete surface whose Gaussian curvature in the optical geometry is $K$, it can be shown that its asymptotic deflection angle $\hat{\alpha}$ is
\[
\hat{\alpha}=-\iint_{S_\infty} K \rd S,
\]
where the integral is taken over the infinite region $S_\infty$ of the surface bounded by the light ray and excluding the lens. 
This allows an iterative computation of the deflection angle: taking a straight line as the zeroth approximation of the light ray, 
the leading term of $\hat{\alpha}$ can be found easily from this equation. Now the main challenge in applying this method to the Kerr solution is 
the fact that its optical geometry is defined by a Randers manifold, a special type of Finsler manifold. While generalizations of the Gauss-Bonnet 
theorem to Finsler surfaces exist (\cite{bcs00}, p. 106), including surfaces with boundary \cite{iss10}, we shall see that the difficulty of an intrinsically Finslerian description may be partially circumvented by working in a suitable osculating Riemannian manifold, thus reviving an early approach to Finsler geometry by Naz\i m \cite{n36}. In addition to being mathematically simpler, this concept is also well suited to the physical problem: since we are currently interested in small deflection angles only, the relevant deviation of the optical Randers geometry from a Riemannian geometry is also small, so using an osculating Riemannian manifold seems natural.

The outline of this paper is therefore as follows. In order to give a mostly self-contained description, we will begin in Section \ref{sec-kerr-randers} 
by briefly sketching some fundamental concepts of Finsler geometry which will be used. Further details can be found, for example, in the monograph by Bao, 
Chern and Shen \cite{bcs00}. This section also defines the optical geometry of the Kerr solution and its Randers structure. Section \ref{sec-nazim} deals with Naz\i m's construction of a Riemannian manifold osculating a Finsler manifold, showing how this notion a can be adapted to the present problem of light deflection in the optical Randers geometry of the equatorial plane in the Kerr solution. Then this idea is applied in Section \ref{sec-lensing} to demonstrate how the two leading terms of the asymptotic gravitational lensing deflection angle of order $\mathcal{O}(m)$ and $\mathcal{O}(ma)$ in the mass and angular momentum parameters can be derived thus, followed by concluding remarks on the benefit of this method in Section \ref{sec-conclusion}. Latin indices are used throughout to denote spatial coordinates, and it will be indicated whether a 3-dimensional or 2-dimensional manifold is considered. Vector and tensor components, thus labelled, 
are with respect to the corresponding coordinate induced basis.

\section{Kerr-Randers optical geometry}
\label{sec-kerr-randers}
A Finsler metric on a smooth manifold $M$ (possibly with boundary) can be defined briefly as a smooth, real, nonnegative function $F(x,X)$ of $x\in M$ 
and $X \in T_xM$ which is positively homogeneous of degree one in $X$, and convex in the sense that the Hessian
\begin{equation}
g_{ij}(x,X)=\frac{1}{2}\frac{\partial^2F^2(x,X)}{\partial X^i \partial X^j}
\label{hessian}
\end{equation}
is positive definite. Then $F^2(x,X)=g_{ij}(x,X)X^iX^j$ from homogeneity, so that $F$ is positive for nonzero vectors and becomes a 
Riemannian norm if the Hessian $g_{ij}$ is independent of vectors which, in this case, becomes a Riemannian metric itself. Hence it is useful to 
characterize the deviation of the Finslerian from the Riemannian case by the symmetric Cartan tensor (see, e.g., \cite{bcs00}, p. 22),
\begin{equation}
C_{ijk}(x,X)=\frac{1}{2}\frac{\partial g_{ij}(x,X)}{\partial X^k}.
\label{cartan}
\end{equation}
Notice also that the Hessian (\ref{hessian}) is positively homogeneous of degree zero. Furthermore, one can construct its inverse $g^{ij}$ in the 
sense that $g_{ij}(x,X)g^{jk}(x,v)=\delta^k_i$, where the dual $v$ of $X$ is defined by $v_i=g_{ij}(x,X)X^j$, and use it to write down formal Christoffel symbols (c.f. \cite{bcs00}, p. 33),
\[
 \Gamma^i_{jk}(x,X)=\frac{1}{2}g^{il}(x,v)\left(\frac{\partial g_{lj}(x,X)}{\partial x^k}+\frac{\partial g_{lk}(x,X)}{\partial x^j}
 -\frac{\partial g_{jk}(x,X)}{\partial x^l}\right),
\]
which are not connection components in general, but clearly reduce to components of the Levi-Civita connection compatible with $g_{ij}$ 
if $F$ is Riemannian. Considering curves $\gamma: t\mapsto x(t)\in M$ that are chosen to be arc-length parametrized, 
so that $\rd t= F(x, \rd x)$ and tangent vectors $\dot{x}$ have unit Finsler length, then a geodesic $\gamma_F$ of the Finsler manifold 
$(M,F)$ is a stationary curve from $\delta\int_\gamma F(x,\dot{x}) \rd t=0$ and obeys (see, e.g., \cite{bcs00}, p. 125)
\begin{equation}
\ddot{x}^i+\Gamma^i_{jk}(x,\dot{x})\dot{x}^j\dot{x}^k=0.
\label{geodesic}
\end{equation}
Now, a Randers metric is a special Finsler metric of the form
\begin{equation}
F(x,X)=\sqrt{a_{ij}(x) X^i X^j}+b_i (x)X^i, 
\label{randers}
\end{equation}
where $a_{ij}$ denotes a Riemannian metric and $b_i$ a one-form satisfying the condition $a^{ij}b_ib_j<1$ in $M$, which ensures both the 
required nonnegativity and convexity of $F$ (see, e.g., \cite{bcs00}, p. 18).

While this structure was first described by Randers \cite{r41} as a model for a spacetime with a time-reversal asymmetry, it naturally arises in the optical geometry of stationary spacetimes, which was originally studied as a useful framework to discuss inertial forces (e.g., \cite{acl88} and \cite{pc90}). To see how the Randers metric arises in the case of the Kerr solution, consider the line element of the Kerr spacetime in the Boyer-Lindquist coordinates $(t,r,\vartheta,\varphi)$,
\begin{equation}
\rd s^2=-\frac{\Delta}{\rho^2}(\rd t-a\sin^2\vartheta \rd \varphi)^2+\frac{\sin^2\vartheta}{\rho^2}((r^2+a^2)\rd \varphi-a\rd t)^2+\frac{\rho^2}{\Delta}\rd r^2+\rho^2\rd \vartheta^2,
\label{kerr}
\end{equation}
defining, as usual, 
\[
\Delta= r^2-2mr+a^2, \quad \rho^2=r^2+a^2\cos^2 \vartheta,
\]
with the ADM mass $m$ and the angular momentum parameter $a=J/m$ with ADM angular momentum $J$. Regarding null curves in the Kerr spacetime (\ref{kerr}), one finds that the coordinate time along their spatial projections obeys a Randers metric of the form (\ref{randers}) with 
$\rd t=F\left(x,\rd x\right)$, where
\begin{align}
a_{ij}\rd x^i\rd x^j&= \frac{\rho^4}{\Delta-a^2\sin^2\vartheta}\left(\frac{\rd r^2}{\Delta}+\rd \vartheta^2+ \frac{\Delta \sin^2 \vartheta}{\Delta-a^2\sin^2\vartheta}\rd \varphi^2\right), \label{a} \\
b_i\rd x^i&= -\frac{2mar\sin^2\vartheta}{\Delta-a^2\sin^2\vartheta}\rd \varphi, 
\label{b}
\end{align}
and the condition for nonnegativity and convexity of $F$ translates to the requirement that 
\[
\frac{(2mar\sin\vartheta)^2}{\Delta\rho^4}<1,
\]
which holds precisely outside the ergoregion of the Kerr solution. A more detailed discussion of this and further examples can be found in \cite{ghww09}. 
Then, by Fermat's principle in general relativity (see Perlick \cite{p00} for a discussion of the general case and applications), spatial light rays are curves which, starting at a fixed light emission event, are selected by stationary arrival time at the observer. So for a stationary observer in the Kerr spacetime, a spatial light ray is stationary in the sense that
\[
0=\delta \int_\gamma \rd t=\delta \int_\gamma F\left(x,\dot{x}\right) \rd t,
\]
so that it is in fact a geodesic $\gamma_F$ satisfying equation (\ref{geodesic}) with respect to the Randers metric given by equations (\ref{a}) and (\ref{b}). It is therefore convenient to define the optical geometry of the Kerr solution as the spatial 3-manifold outside the ergoregion equipped with this Randers metric $F$, whose geodesics $\gamma_F$ are the spatial light rays.

Henceforth, however, we shall specialize the discussion to the equatorial plane $\vartheta=\pi/2$ of the Kerr solution which, by symmetry, is geodesically complete so that light rays in this plane remain therein. The corresponding optical geometry is then given by the 2-manifold $M$ of the equatorial plane outside the ergoregion, described by the $(r,\varphi)$-coordinates with $r>2m$, and the Randers metric
\begin{equation}
 F\left(r,\varphi,\frac{\rd r}{\rd t},\frac{\rd \varphi}{\rd t}\right)=\sqrt{\frac{r^4}{\Delta(\Delta-a^2)}\left(\frac{\rd r}{\rd t}\right)^2+\frac{r^4\Delta}{(\Delta-a^2)^2}\left(\frac{\rd \varphi}{\rd t}\right)^2}-\frac{2mar}{\Delta-a^2}\frac{\rd \varphi}{\rd t},
\label{equatorial}
\end{equation}
from equations (\ref{randers}), (\ref{a}) and (\ref{b}).

\section{Adapting Naz\i m's construction}
Applying a notion first introduced by Naz\i m (\cite{n36}, p. 22), we shall now proceed with the construction of a Riemannian manifold $(M,\gb)$ osculating the Randers manifold $(M,F)$. Its Riemannian metric $\gb$ is obtained from the Hessian (\ref{hessian}) by choosing a smooth nonzero vector field $\xb$ over $M$ that contains the tangent vectors along the geodesic $\gamma_F$, that is $\xb(\gamma_F)=\dot{x}$, defining
\begin{equation}
\gb_{ij}(x)=g_{ij}\left(x,\xb(x)\right),
\label{metric}
\end{equation}
whose Levi-Civita connection is denoted by $\bar{\Gamma}^i_{jk}$. Geometrical quantities in the osculating Riemannian manifold, for instance angles between vectors in $T_xM$ measured by the metric $\gb$, naturally depend on the particular choice of $\xb$. Crucially, however, the geodesic $\gamma_F$ of $(M,F)$ is also a geodesic $\gamma_{\gb}$ of $(M,\gb)$ by construction. This can be seen easily from the partial derivatives of $\gb_{ij}$ (c.f. \cite{n36}, p. 24),
\begin{align*}
 \frac{\partial \gb_{ij}(x)}{\partial x^k}&=\frac{\partial g_{ij}\left(x,\xb\right)}{\partial x^k}+\frac{\partial g_{ij}\left(x,\xb\right)}{\partial \xb^l}\frac{\partial \xb^l(x)}{\partial x^k}\\
 &=\frac{\partial g_{ij}\left(x,\xb\right)}{\partial x^k}+2C_{ijl}\left(x,\xb(x)\right)\frac{\partial \xb^l(x)}{\partial x^k}
\end{align*}
using (\ref{metric}) and (\ref{cartan}). Computing $\Gamma^i_{jk}(x,\dot{x})$ and $\bar{\Gamma}^i_{jk}(x)$ along the geodesic $\gamma_F$ given by equation (\ref{geodesic}), one obtains
\[
0=\ddot{x}^i+\Gamma^i_{jk}(x,\dot{x})\dot{x}^j\dot{x}^k=\ddot{x}^i+\bar{\Gamma}^i_{jk}(x)\dot{x}^j\dot{x}^k,
\]
since terms of the form $C_{ijk}(x,\dot{x})\dot{x}^i$ vanish due to the homogeneity of the Hessian, so that $\gamma_F=\gamma_{\gb}$ as required. Angles defined along the geodesic in $(M,\gb)$ can thus be considered intrinsic to $(M,F)$ itself. We can therefore work with an osculating Riemannian manifold derived from the Randers optical geometry to compute the deflection angle of light rays in the equatorial plane of the Kerr solution.

\begin{figure}
\centering
\includegraphics{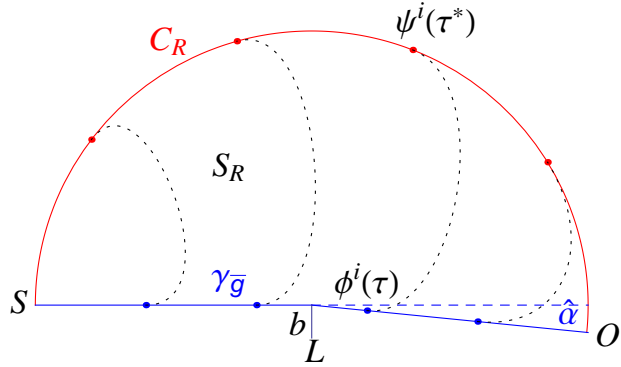}
\caption{Naz\i m's construction adapted to gravitational lensing by the Kerr black hole, at $L$ in the equatorial $(r,\varphi)$-plane, deflecting a light ray $\gamma_{\gb}$ from the light source $S$ to the observer $O$ by the asymptotic angle $\hat{\alpha}$ relative to the undeflected ray (dashed line). The dotted curves show $x^i(\sigma,\tau)$ defined in equation $(\ref{nazim2})$ with $y^i(\sigma,\tau)=0$ for four values of $\tau$.}
\label{fig-nazim}
\end{figure}

Starting with the Randers optical geometry $(M,F)$, then, consider a region $S_R\subset M$ in the equatorial plane bounded by the light ray $\gamma_F$ passing the Kerr lens situated at the origin at coordinate-radius $b$, and a circle segment $C_R$ of coordinate-radius $R$ centered at the lens which intersects $\gamma_F$ in the points $S$ and $O$. We shall assume $R\gg b\gg m$ so that, physically, the deflection angle of the light ray is small and $S$ and $O$ correspond to a stationary light source and observer in the asymptotically Euclidean region. Suppose that the two boundary curves are given by
\begin{align*}
\gamma_F: \quad x^i(t)&=\phi^i(t), \quad 0\leq t \leq l, \\
C_R: \quad x^i(t)&=\psi^i(t), \quad 0\leq t \leq l^\ast,
\end{align*}
where $\phi^i(0)=\psi^i(l^\ast)$ at $S$ and $\phi^i(l)=\psi^i(0)$ at $O$, noting that the lengths $l$ of $\gamma_F$ and $l^\ast$ of $C_R$ between $S$ and $O$ are with respect to $F$ and therefore depend on the direction of the curves indicated. Now letting $\tau=t/l\in(0,1)$ along $\gamma_F$ and $\tau^\ast=1-t/l^\ast\in(0,1)$ along $C_R$, we can pair each point $\phi^i(\tau)$ on $\gamma_F$ with precisely one point $\psi^i(\tau^\ast)$ on $C_R$ by taking $\tau=\tau^\ast$. Next, one can construct a family of smooth curves $x^i(\sigma,\tau)$, where each curve is parametrized by $\sigma \in [0,1]$ and labelled by $\tau$, such that for each point pair there is precisely one curve which touches the boundary curves $\gamma_F$ and $C_R$,
\begin{equation}
\begin{split}
x^i(0,\tau)&=\phi^i(\tau),\quad \ \frac{\rd x^i}{\rd \sigma}(0,\tau)=\frac{\rd \phi^i}{\rd t}(\tau):=\dot{\phi}^i(\tau),\\
x^i(1,\tau)&=\psi^i(\tau),\quad \frac{\rd x^i}{\rd \sigma}(1,\tau)=\frac{\rd \psi^i}{\rd t}(\tau):=\dot{\psi}^i(\tau).
\end{split}
\label{nazim1}
\end{equation}
Then the tangent vector field of this family of curves is a suitable smooth nonzero vector field
\begin{equation}
\xb^i\left(x(\sigma,\tau)\right)=\frac{\rd x^i}{\rd \sigma}(\sigma,\tau)
\label{field}
\end{equation}
which can be used to define a Riemannian metric $\gb_{ij}$ according to (\ref{metric}). A family of curves that satisfies condition (\ref{nazim1}) can 
be written down explicitly as follows (c.f. \cite{n48}),
\begin{equation}
\begin{split}
x^i(\sigma,\tau)&=\phi^i(\tau)+\dot{\phi}^i(\tau)\sigma+\left(3\psi^i(\tau)-3\phi^i(\tau)-\dot{\psi}^i(\tau)
-2\dot{\phi}^i(\tau)\right)\sigma^2  \\
&\quad+\left(2\phi^i(\tau)-2\psi^i(\tau)+\dot{\psi}^i(\tau)+\dot{\phi}^i(\tau)\right)\sigma^3
+y^i(\sigma,\tau)\sigma^2(1-\sigma)^2,
\end{split}
\label{nazim2}
\end{equation}
where the functions $y^i(\sigma,\tau)$ are smooth but otherwise arbitrary, and are chosen to make $\xb$ nonzero over $S_R$. At the points $S$ and $O$, $\xb$ and hence $\gb$ is not defined by this construction but $\gb$ is asymptotically Euclidean by assumption. Hence, $(S_R,\gb)$ with the metric given by equations (\ref{nazim2}), (\ref{field}) and (\ref{metric}) shall be the osculating Riemannian manifold with boundary applied to our gravitational lensing problem, which is sketched in Fig. \ref{fig-nazim}. As discussed above, the light ray $\gamma_F$ is then also a geodesic $\gamma_{\gb}$ in $(S_R,\gb)$. 
\label{sec-nazim}

\section{Application to gravitational lensing}
\label{sec-lensing}
Having constructed the osculating Riemannian manifold $(S_R,\gb)$ with boundary $\partial S_R=\gamma_{\gb} \cup C_R$ as described in the previous section, we can now apply the Gauss-Bonnet method \cite{gw08} to evaluate the asymptotic light deflection angle. The Gauss-Bonnet theorem applied to the region $S_R$ in $M$ states that
\begin{equation}
\iint_{S_R} K \rd S+\oint_{\partial S_R}\kappa\rd t=2\pi\chi(S_R)-(\theta_O+\theta_S),
\label{gaussbonnet}
\end{equation}
where the Gaussian curvature $K$, the geodesic curvature $\kappa$, and the exterior jump angles $\theta_S$ and $\theta_O$ at the vertices $S$ and $O$ of the boundary curve, respectively, are understood to be with respect to the Riemannian metric $\gb$. Again, since $S$ and $O$ are assumed to be in the asymptotically Euclidean region, these jump angles are well-defined at the vertices. Now we orient the coordinate system so that $\varphi(S)=0$ as $R\rightarrow \infty$. Clearly, in this limit, both jump angles become $\pi/2$ so that the term $\theta_O+\theta_S\rightarrow \pi$ and, because the Euler characteristic $\chi(S_R)=1$, the right-hand side of equation (\ref{gaussbonnet}) tends to $\pi$ overall. Furthermore, since $\gamma_{\gb}$ is a geodesic $\kappa(\gamma_{\gb})=0$. Finally, as $R\rightarrow\infty$ we have $\kappa(C_R)\rightarrow R^{-1}$ and also $\rd t\rightarrow R \rd \varphi$ by arc-length parametrization, so the left-hand side of (\ref{gaussbonnet}) becomes
\[
\iint_{S_R} K \rd S+\int_{C_R} \kappa \rd t\stackrel{R\rightarrow \infty}{=}\iint_{S_\infty} K \rd S+\int_0^{\pi+\hat{\alpha}} \rd \varphi,
\]
with the asymptotic deflection angle $\hat{\alpha}$, as shown in Fig. \ref{fig-nazim}. Hence we find, formally, the same expression for $\hat{\alpha}$ as 
mentioned in the Introduction,
\begin{equation}
\hat{\alpha}=-\iint_{S_\infty} K\rd S,
\label{deflection}
\end{equation}
using, however, an osculating Riemannian manifold $(S_R,\gb)$ for our lensing problem in the Finslerian optical geometry. The asymptotic deflection angle can then be computed iteratively, starting with an undeflected line. To illustrate this method, we shall now work out the leading terms of the deflection angle in $m$ and $a$ from (\ref{deflection}), taking the line $r(\varphi)=b/\sin\varphi$ as the zeroth approximation of the deflected light ray $\gamma_{\gb}$ and inner boundary of the integration domain in (\ref{deflection}). For this purpose, it suffices to use only the leading terms of the vector field $\xb=(\xb^r,\xb^\varphi)(r,\varphi)$ from (\ref{field}) near this boundary,
\[
\xb^r=\frac{\rd r}{\rd t}=-\cos\varphi+\mathcal{O}(m,a), \quad \xb^\varphi=\frac{\rd \varphi}{\rd t}=\frac{\sin^2\varphi}{b}+\mathcal{O}(m,a),
\]
in the Randers metric (\ref{equatorial}). Hence the metric $\gb$ of the osculating Riemannian manifold can be found from (\ref{metric}), whose components are
\begin{align*}
\gb_{rr}&= 1+\frac{4m}{r}-\frac{2mar}{b^3}\frac{\sin^6\varphi}{\left(\cos^2\varphi+\frac{r^2}{b^2}\sin^4\varphi\right)^\frac{3}{2}} +\mathcal{O}(m^2,a^2),\\
\gb_{r\varphi}&=\frac{2ma}{r}\frac{\cos^3\varphi}{\left(\cos^2\varphi+\frac{r^2}{b^2}\sin^4\varphi\right)^\frac{3}{2}}+\mathcal{O}(m^2,a^2), \\
\gb_{\varphi\varphi}&=r^2+2mr-\frac{2mar}{b}\frac{\sin^2\varphi\left( 3\cos^2\varphi+2\frac{r^2}{b^2}\sin^4\varphi\right)}{\left(\cos^2\varphi+\frac{r^2}{b^2}\sin^4\varphi\right)^\frac{3}{2}}+\mathcal{O}(m^2,a^2).
\end{align*}
The determinant of this metric is given by,
\begin{equation}
\det \gb=r^2+6mr-\frac{6mar}{b}\frac{\sin^2\varphi}{\sqrt{\cos^2\varphi+\frac{r^2}{b^2}\sin^4\varphi}}+\mathcal{O}(m^2,a^2).
\label{det}
\end{equation}
One can now compute the corresponding Gaussian curvature of the surface,
\begin{equation}
\begin{split}
K&=\frac{\bar{R}_{r\varphi r\varphi}}{\det \gb}=\frac{1}{\sqrt{\det \gb}}\left[\frac{\partial}{\partial \varphi}\left(\frac{\sqrt{\det \gb}}{\gb_{rr}}\bar{\Gamma}^\varphi_{rr} \right)-\frac{\partial}{\partial r}
\left(\frac{\sqrt{\det \gb}}{\gb_{rr}}\bar{\Gamma}^\varphi_{r\varphi}\right)\right] \\
&=-\frac{2m}{r^3}+\frac{3ma}{b^2r^2}\f+\mathcal{O}(m^2,a^2),
\end{split}
\label{gauss}
\end{equation}
where
\[
\begin{split}
\f&=\frac{\sin^3\varphi}{\left(\cos^2\varphi+\frac{r^2}{b^2}\sin^4\varphi\right)^\frac{7}{2}}\left[2\cos^6\varphi\left(-2+5\frac{r}{b}\sin\varphi\right)+\cos^4\varphi\sin^2\varphi\left(-2+9\frac{r}{b}\sin\varphi-10\frac{r^3}{b^3}\sin^3\varphi\right)\right. \\
&\left.+2\frac{r}{b}\cos^2\varphi\sin^5\varphi\left(2-\frac{r^2}{b^2}+\frac{r^2}{b^2}\cos2\varphi+4\frac{r}{b}\sin\varphi\right)+\frac{r^2}{b^2}\left(-\frac{r}{b}\sin^9\varphi+2\frac{r^3}{b^3}\sin^{11}\varphi+\sin^4 2\varphi\right)\right].
\end{split}
\]
Approximating the integration domain of (\ref{deflection}) as described above and using $K$ from (\ref{gauss}), we obtain
\begin{equation}
\hat{\alpha}\simeq-\int_0^\pi \int_\frac{b}{\sin\varphi}^\infty K \sqrt{\det \gb} \rd r \rd \varphi\simeq -\int_0^\pi \int_\frac{b}{\sin \varphi}^\infty 
\left(- \frac{2m}{r^2} +\frac{3ma}{b^2r}\f \right)\rd r \rd \varphi.
\label{deflection1}
\end{equation}
Of course, our lensing setup in the given coordinate system supposes that the light ray is a prograde orbit relative to the rotation of the Kerr black hole. Taking this into account and integrating,
\[
\int_0^\pi\int_\frac{b}{\sin\varphi}^\infty  \frac{2m}{r^2}\rd r \rd \varphi=\frac{4m}{b}, \quad \int_0^\pi\int_\frac{b}{\sin\varphi}^\infty \frac{3ma}{b^2r}\f \rd r  \rd \varphi=\frac{4ma}{b^2},
\]
one indeed recovers from (\ref{deflection1}) the expected leading terms in $m$ and $a$ of the asymptotic deflection angle (first found for the Kerr solution in \cite{bl67}),
\begin{equation}
\hat{\alpha}\simeq\frac{4m}{b}\pm\frac{4ma}{b^2},
\label{deflection2}
\end{equation}
where the positive and negative sign is for a retrograde and prograde light ray, respectively.

\section{Concluding remarks}
\label{sec-conclusion}
While the leading terms of the Kerr deflection angle (\ref{deflection2}) are, of course, well-established, the main interest of the method proposed in this article is conceptual.

Firstly, even for small deflection angles, optical geometry offers a more geometrical perspective on gravitational lensing compared to the standard quasi-Newtonian thin-lens approximation. Whereas the standard approximation uses a notional flat space where the deflection of light is put in by hand using an impulse calculated separately, in optical geometry light rays are treated more naturally as geodesics of a curved space.  But rather than computing null geodesics in spacetime itself, optical geometry considers spatial light rays only so, in this simpler framework, physical lens models can be implemented easily (for other examples, see \cite{gw08}).

Secondly, from this point of view, one obtains the conceptually rather surprising result that the deflection of a light ray is determined by a quantity \textit{outside} of itself relative to the lens, according to equation (\ref{deflection}). This reflects the fact that gravitational lensing can be regarded as a global effect: indeed, topological conditions for the existence of more than one light ray joining source and observer, that is, the occurrence of multiple images, can be exhibited using (\ref{gaussbonnet}) as shown in \cite{gw08}. A detailed recent study of topological lensing effects in this context can be found in \cite{cgs11}.

Thirdly, this work contributes to the physical applications of Finsler geometry. In particular, the osculating Riemannian approach promoted here may be useful for other physical problems which are properly described by Finsler geometry but are, in effect, nearly Riemannian. For instance, this could be applied to particle motion in a magnetic field. Of course, it would also be interesting to see a treatment of gravitational lensing in the Randers optical geometry which makes use of a formulation intrinsic to the Finsler geometry itself, for example along the lines of \cite{iss10}. This would allow an extension of the present work beyond the assumption that both source and observer lie in the asymptotically Euclidean region, including a more generally applicable treatment of jump angles at the vertices, which would be relevant for studies of stronger deflection closer to the Kerr black hole. 

\acknowledgement
This work was supported by World Premier International Research Center Initiative (WPI Initiative), MEXT, Japan.

\end{document}